# Achieve equilibrium outside the contact angle hysteresis

Lei Liu[1,3,4], Guanlong Guo[3,4], Kaiyu Wang[1,3,4], Linke Chen[3,4], Yangyang Fan[4,5], Sergio Andres Galindo Torres[3,4], Jiu-an Lv[4,5], Herbert E. Huppert[6], Changfu Wei[7], Liang Lei[2,3,4*]

[1]College of Environmental and Resources Science, Zhejiang University, Hangzhou, Zhejiang 310058, China

[2]Research Center for Industries of the Future, Westlake University, Hangzhou, Zhejiang 310030, China

[3]Key Laboratory of Coastal Environment and Resources of Zhejiang Province, School of Engineering, Westlake University, Hangzhou, Zhejiang310030, China

[4]Institute of Advanced Technology, Westlake Institute for Advanced Study, Hangzhou, Zhejiang 310024, China

[5]Key Laboratory of 3D Micro/Nano Fabrication and Characterization of Zhejiang Province, School of Engineering, Westlake University, Hangzhou 310024Zhejiang Province, China;

[6]Institute of Theoretical Geophysics, King's College, University of Cambridge, King's Parade, Cambridge CB2 1ST, United Kingdom;

[7]State Key Laboratory of Geomechanics and Geotechnical Engineering, Institute of Rock and Soil Mechanics, Chinese Academy of Sciences, Wuhan 43001, Hubei, China

*Corresponding author: Liang Lei leiliang@westlake.edu.cn

## Abstract

It is common belief that the equilibrium contact angle, corresponding to the minimum system energy state, lies between advancing and receding contact angles. Here, we derive advancing and receding contact angles considering the micro contacting processes on ideal rough 2D surfaces. Equilibrium contact angles obtained via energy minimization can be smaller than the receding contact angle and reach 0 degrees, at which hysteresis diminishes on the super-hydrophilic surface. Gibbs free energy analyses, numerical simulations and physical experiments all confirm these new findings.

Contact angle is a fundamental concept in interfacial science. Young's equation provides the first understanding of contact angle based on mechanical equilibrium [1]. However, the actual contact angle varies during droplet advancing and receding [2–5]. The hysteresis, that is, the difference between critical advancing and receding contact angles, affects printing, painting, coating, fogging, distribution of herbicides and insecticides, and multiphase flow in porous media [6–8]. Surface roughness/microstructure, surface heterogeneity, solid surface deformation, and adaptation contribute to contact angle hysteresis [9–15]. It is generally believed that the equilibrium contact angle $\theta_e$, determined by the minimization of the Gibbs free interfacial energy, lies somewhere between the advancing and receding contact angles [9]. Here we prove both theoretically and experimentally that the equilibrium contact angle can be outside this hysteresis.

**Contact angle hysteresis derivation.** Consider a 2D case with a unit length in the third dimension where a water droplet sits on a 1D rough surface with cylindrical humps (Fig.1). Ignore gravity and consider the droplet shape to be a circle. The intrinsic contact angle (assumed only dependent on the chemistry of fluids and solid surface) $\theta$ is universal with no hysteresis. Interface curvature is uniform due to the identical pressure difference between liquid and gas. During advancing (Fig.1a), the common point where solid-liquid-gas joins slides down along the hump surface until the liquid-gas interface touches the next hump, and a Haines jump (a sudden jump of the fluid interface accompanied by fluid redistribution and a transient pressure response) [16] occurs and the system reaches a new static state. Two cases could occur during receding: (1) the liquid-gas interface touches the next hump (Fig.1b); and (2) the common point on the same hump meets (Fig.1c). Residual water appears between humps in the first case, while the water droplet pulls all water towards the next hump in the second case. See mirror cases in the supplementary material.

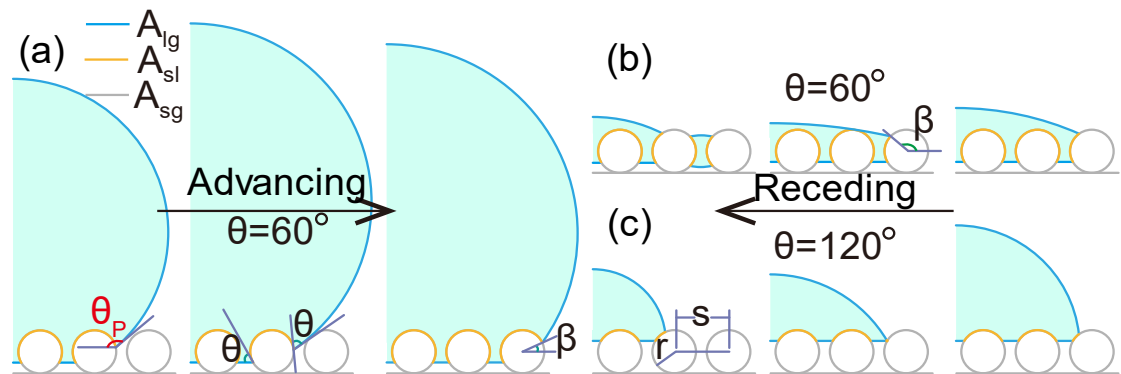

Figure 1. Droplets advancing (a, from left to right) and receding (b and c, from right to left) on rough surfaces with cylindrical humps. $A_{lg}$, $A_{sl}$, $A_{sg}$ are the liquid-gas, solid-liquid and solid-gas interfacial areas. Haines jumps occur when different interfaces touch each other.

Straightforward geometric analysis enables the calculation of the apparent contact angle $\theta_p$ and droplet volume $V_d$ (cyan domain) as functions of $\theta$, contacted humps number $n$, and angle $\beta$ determining the margin common point (the outermost contact point between droplet and cylindrical hump), hump radius $r$ and center distance between two adjacent humps $s$. An implicit relation is established between the apparent contact angle $\theta_p$ and droplet volume $V_d$. Besides, the asymptotes of advancing, $\theta_a$, Eq. (1), and receding contact angles, $\theta_{r1}$, Eq. (2), and $\theta_{r2}$, Eq. (3), are derived when $n$ approaches infinity. The derivation is detailed in the supplementary material.

$$\sin(\pi - \theta_a) = r(1 + \cos\theta)/s \qquad (1)$$

$$\sin\theta_{r1} = r(1 - \cos\theta)/s \qquad (2)$$

$$\theta_{r2} = 2\theta - \pi \qquad (3)$$

**Energy analyses and the equilibrium contact angle.** The Gibbs free interfacial energy, $G$, is defined as the difference between the final state and initial state with a dry solid surface and a cylindrical water droplet [17,18].

$$G = A_{sl}(\gamma_{sl} - \gamma_{sg}) + (A_{lg} - A_0)\gamma_{lg}, \qquad (4)$$

where $\gamma_{sl}$, $\gamma_{sg}$, $\gamma_{lg}$ are the solid-liquid, solid-gas and liquid-gas interfacial tensions. $A_{sl}$ and $A_{lg}$ are the solid-liquid and liquid-gas interfacial areas (Fig. 1). $A_0 = 2\pi\sqrt{V_d/\pi}$ is the surface area of the initial cylindrical water droplet. The equilibrium contact angle is obtained by minimizing $G$. The derivation is detailed in the supplementary material.

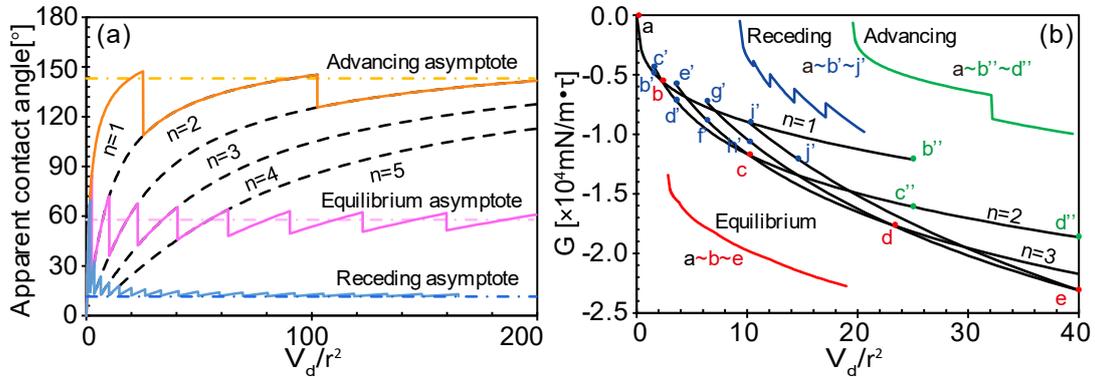

Figure 2. Paths of (a) Apparent contact angles and (b) Gibbs free interfacial energy versus droplet volume with different $n$ (contacted humps number) at advancing, receding and the equilibrium states.

**Contact angle oscillation around the predicted results.** We plot the apparent contact angle versus the droplet volume normalized by a unit volume $r^2$ (Fig. 2a), where $\theta$ is 60°, and $s$ equals $2.5r$. The normalization makes results scale independent. Black dash lines denote the apparent contact angle-volume relationships when $n$ varies from 1 to 5. The orange and blue solid lines represent the droplet advancing and receding processes to next humps (Fig. 2a). The advancing and receding angles oscillate around the asymptote values predicted by Eqs. (1) and (2), with a decreasing oscillation amplitude as $n$ increases. The receding contact angle $\theta_{r2}$ is seemingly independent of the surface structure dimension. Yet case (2) only occurs with a combination of large $\theta$ and small $s/r$. Numerical simulations (details in supplementary material) using the pseudopotential multicomponent Lattice Boltzmann method (LBM) validate our thought analysis.

$G$ drops upon the Haines jump, during both advancing and receding (green and blue lines in Fig. 2b). This energy cliff, which would induce droplet vibration and gets dissipated via fluid internal friction. $\theta_e$ (purple solid line in Fig. 2a), is obtained by energy minimization via variation of contacted hump number with a given droplet volume. Its asymptote (Eq. 5) is given following our previous approach [18]. $\theta_e$ also zigzags around and gradually approaches the predicted asymptote, while the $G$ curve at $\theta_e$ (red line) is continuous but not smooth.

$$cos\theta_p = -1 + 2r[(\pi - \theta)cos\theta + sin\theta]/s. \qquad (5)$$

**Experimental validation of Contact angle hysteresis.** Quasi-2D physical experiments validate these analytical results (Details in supplementary material). The rough surface is made of fishing wires wrapped around designed grooves (Fig. 4a). We use the intrinsic contact angle ($\theta=71°$) of the fishing wire to calculate the apparent advancing and receding contact angles of the rough surface. Geometrical parameters are: $s = 0.25$ mm, $r = 0.0625$ mm, and $s/r = 4$. The measured advancing contact angle is 160.4° (by instrument OCA25), matching well with the predicted $\theta_a = 160.5°$. Measured receding contact angles range from 26.7° to 65.4°, which are significantly larger than the predicted $\theta_r = 9.7°$. The primary reason is that the advancing of the actual 3D water droplet approximates the 2D assumptions in the earlier analyses, while the receding is obviously affected by the 3D droplet shape because the contact line on the margin fishing wire approaches to a point before the Haines jump (See videos 4, 5 and 6 for the dynamic processes).

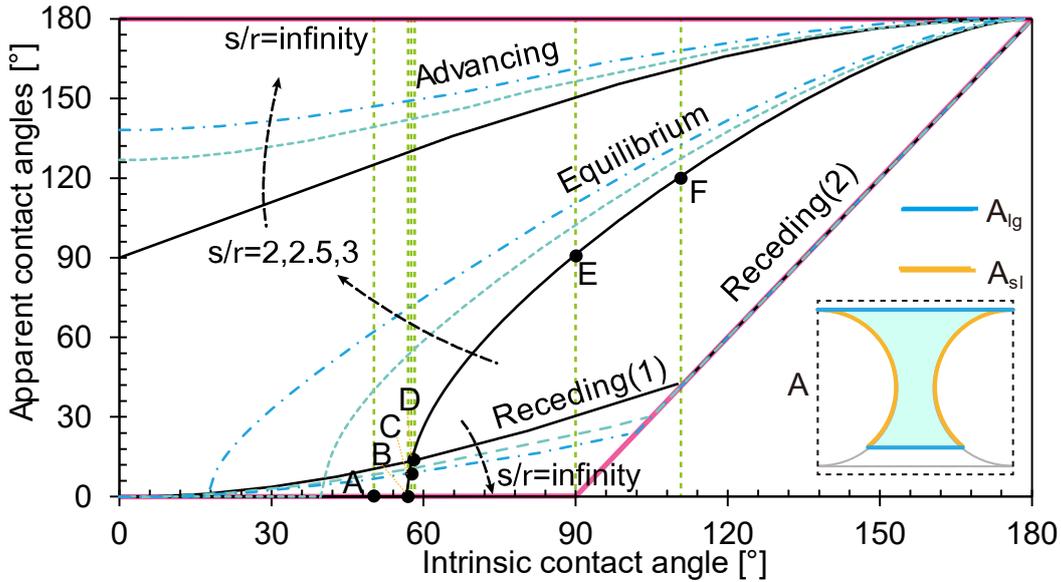

Figure 3. Apparent contact angle variation with intrinsic contact angle and different ratios of $s/r$ = 2, 2.5, 3. The red solid lines are the upper and lower bounds of advancing and receding contact angles when $s/r$ reaches infinity. Points A, B, C, D, E,

and F on the equilibrium contact angle line at $s/r = 2$ are selected for further analyses on $G$ (Figure 5). The inset demonstrates the liquid distribution in one unit at point A and the corresponding interfacial areas $A_{lg}$ and $A_{sl}$.

**Predicted contact angles on different surfaces.** $\theta_a$, $\theta_e$, $\theta_r$ are plotted together as functions of $\theta$ and the structure parameter $s/r$ (Fig. 3). Contact angle hysteresis increases with $s/r$. When $s/r$ increases to infinity, the upper limit for $\theta_a$ is 180°, independent of $\theta$. However, the lower limits for $\theta_r$ are 0° with an infinite $s/r$ in case 1 and $2\theta - \pi$ in case 2 (red solid line). The transition between these two cases is the joint point of the two curves in Fig. 3. If the material itself is superhydrophobic, $\theta = \pi$, both the advancing and receding contact angles are 180°, showing zero hysteresis.

**A more stable state outside hysteresis.** All current available models predict that $\theta_e$ should be between $\theta_a$ and $\theta_r$. It is bizarre that there are cases where $\theta_e$ is lower than $\theta_r$ (points A, B and C). The underlying physics is that if we can spread the water on a solid surface, the system is more stable, because $G$ at point A with $\theta_e = 0°$ (the inset of Fig. 3) is smaller than any other $\theta_p$ at the same droplet volume.

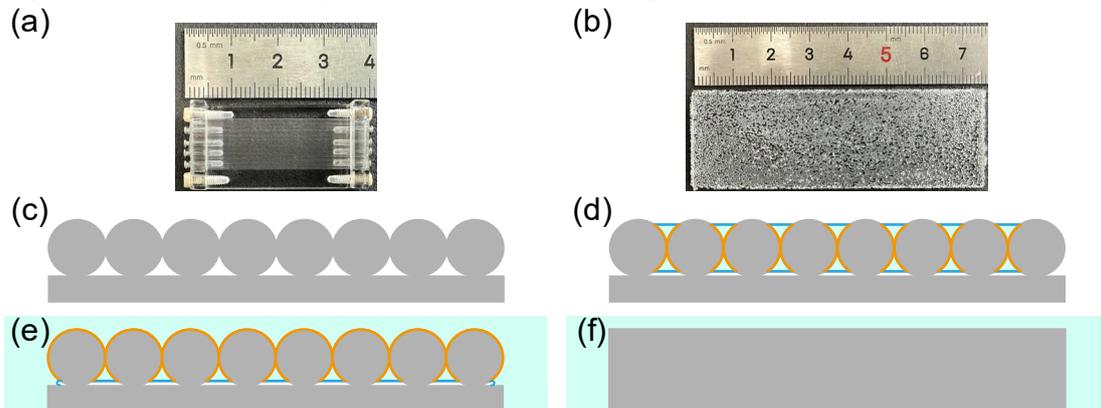

Figure 4. (a) and (b) The rough surfaces made of fish wire and glass beads. (c)-(f) are schematics for surfaces under different conditions. (c) Dry rough surface. (d) Prewetted rough surface with trapped bubble. (e) Prewetted rough surface with trapped gas bubbles submerged in water. (f) Smooth surface submerged in water.

**Experimental validation of the new equilibrium outside the hysteresis.** Another series of experiments were conducted to validate this counterintuitive result (videos in supplementary material). A layer of glass beads glued to a flat slide forms the rough surface (Fig. 4b and 4c). Parameters for this rough surface are average contact angle of glass beads $\theta = 36.14°$, measured by CT images (see image 7 in supplementary material), s = 0.50 mm, $r$ = 0.25 mm, and $s/r$ = 2. The predicted equilibrium contact angle is 0°, calculated by eq (7) in our previous approach [18]. Advancing and receding contact angles are difficult to calculate due to the imperfect geometry and complex 3D interface shape. We choose to physically measure the advancing and receding angles (by instrument OCA25), which are 96.15° and 7.53° (see videos 8 and 9 in supplementary material). Shaking the surface would not produce an apparent contact angle outside the hysteresis. However, if we spread water on the surface first (Fig. 4d), the water droplet spreads quickly upon touching the surface, indicating a 0° contact angle during advancing (video 10 in supplementary material). When the water table (Figs. 4e and 4f) is slightly higher than the smooth and rough surface, by 1mm for example, shaking would expose part of the smooth surface, while the rough surface always remains under a flat water surface (videos 11 and 12 in supplementary material). Note these observations are valid no matter whether the rough surface has gas bubbles trapped or not (Fig. 4e, see video 13 in supplementary material). We therefore conclude that the contact angle on this pre-wetted rough surface is 0°, showing neither advancing nor receding hysteresis.

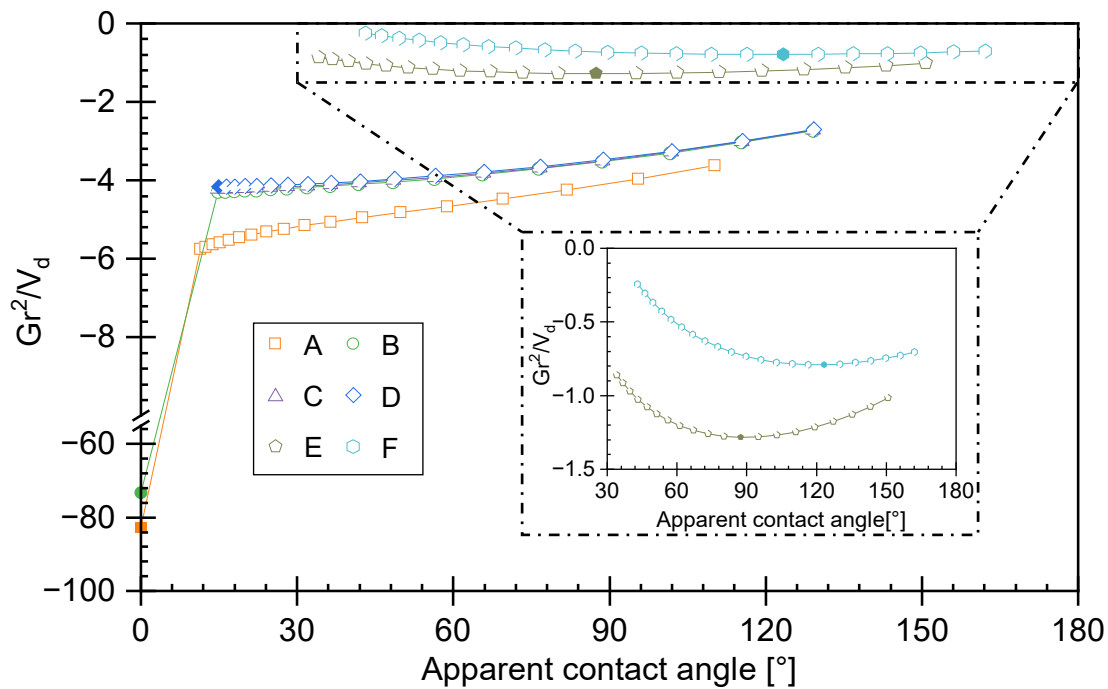

Figure 5. Gibbs free interfacial energy against apparent contact angle at the same droplet volume.

**Explanation of the new equilibrium with energy analyses.** Analyses based on normalized Gibbs free interfacial energy, $Gr^2/V_d$, are conducted to reveal the underlying physics (Figure 5). Six points A, B, C, D, E, and F on the calculated equilibrium contact angle curves (vertical lines in Figure 3) are selected for $G$ comparison with cases as if the same water droplet takes other apparent contact angles. For points E and F, the minimum $G$, that is the equilibrium, is between advancing and receding contact angles. For point D, the equilibrium is the same as $\theta_r$. For points A, B and C, their equilibrium is outside the contact angle hysteresis. These results agree well with our theoretical analysis and experiments.

The case with $s/r = 2$ is widely found in nature, one specific case being the surface of granular material. Daily observation proves the conclusion because there are no water droplets on wet soil, especially when the soil is saturated with water (video 14 in supplementary material).

**Summary.** We show that geometrical constraint is one source of contact angle hysteresis. Theoretical analyses and experimental results confirm that the equilibrium contact angle can be smaller than the receding contact angle. By prefilling the structured surface with water, the contact angle of 0° can be achieved and the contact angle hysteresis disappears. This mechanism can be used to create super-wetting.

## Acknowledgments

This work was supported by the Key Research and Development Program of Zhejiang Province (Grant No. 2023SDXHDX0005), National Natural Science Foundation of China (Grant NO. 42107199) and Research Center for Industries of the Future (RCIF). We thank Chongyan Ma for the help with video filming.

## Author contribution statement

Liang Lei, Lei Liu and Changfu Wei designed the research; Lei Liu and Liang Lei performed the research and analyzed the data; Guanlong Guo and Sergio Andres Galindo Torres contributed to the simulation; Kaiyu Wang, Linke Chen, Yangyang Fan and Jiu-an Lv contributed to the experiments. Lei Liu, Liang Lei and Herbert E. Huppert wrote the paper.

## Supplementary Materials

**Mirror problem**

It is intuitive to consider the advancing of one fluid identical to the receding of the other fluid on the same solid surface. However, there are an odd number of cases as displayed in Fig. 1. Figure S1 displays the one-to-one corresponding mirror problems. The advancing or receding of cyan droplets in Fig. 1 corresponds to the receding and advancing of purple droplets in Fig. S1. Note here we assume the fluid is all magically connected so that they share the same pressure. The key difference between Figs. 1 and S1 is whether the bottom pore space between humps is initially filled with gas or water. Similarly, air bubbles could be trapped during droplet advancing in Fig. S1(b). While the case in Fig. S1(c) is a perfect mirror of Fig. 1(c), i.e., the advancing angle in Fig. S1(c) and receding angle in Fig. 1(c) are supplementary angles; the apparent contact angles in Figs. S1(a) and S1(b) are not the perfectly supplementary angles of those in Figs. 1(a) and 1(b) as shown in the inset in Fig. S1(a). This imperfection decreases when the contacted hump numbers go to infinity and the apparent contact angle approaches its asymptote.

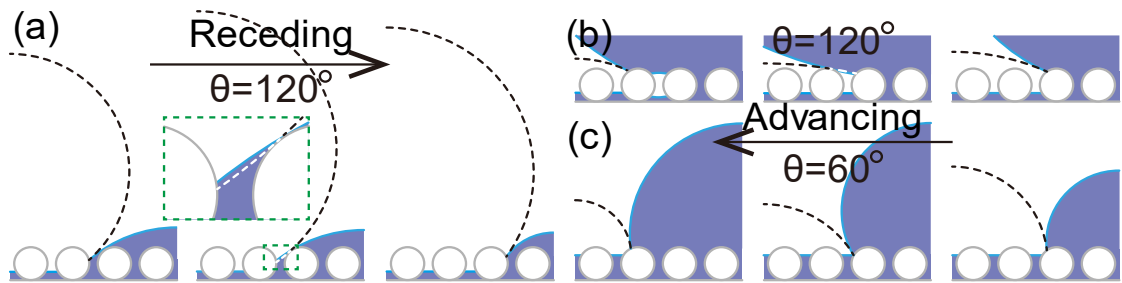

Figure. S1. Mirror cases of Fig. 1.

**The apparent contact angle and radius of the droplet**

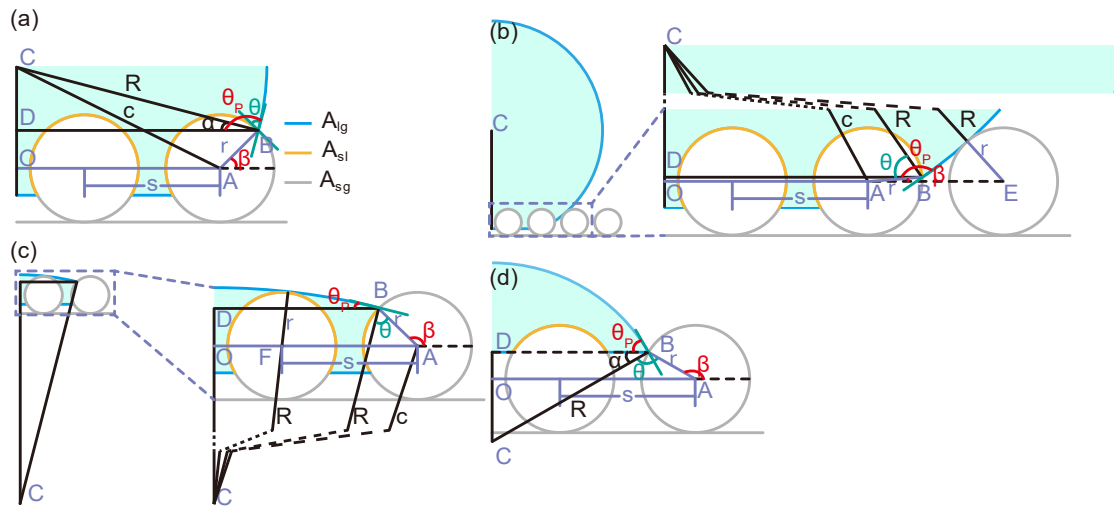

Figure. S2. The advancing and receding processes of a droplet on a rough surface with cylindrical humps. (a) The advancing and receding process. (b) The advancing limit. (c) and (d) the receding limit. Note that the intrinsic contact angle $\theta$ of the droplet is 60° in (a), (b) and (c), and 120° in (d). $R$ and $r$ are the radii of the droplet and cylindrical hump; $\theta_P$ is the apparent contact angle; $s$ is the center distance between two adjacent humps; and $A_{lg}$, $A_{sl}$, and $A_{sg}$ are the surface area of liquid-gas, solid-liquid and solid-gas, respectively.

Assume a droplet with a specific volume is in equilibrium on a flat surface with cylindrical humps, as shown in Fig. S2(a). Define point O as the origin of the coordinate system. The coordinates of point A (the center of cylindrical humps) are $(x_A, 0)$. B is the common point and its position changes with droplet volume and can be determined by $\beta$, the angle between line AB and the extension line OA. Then, the coordinates of point B are $(x_A + r\cos\beta, r\sin\beta)$, where $r$ is the radius of the cylindrical hump. Assume the length of BC is $R$, the radius of the droplet. The coordinates of point C, $(x_A + r\cos\beta - R\cos\alpha, r\sin\beta + R\sin\alpha)$, can be obtained, dependent on the point B and the angle, $\alpha = \theta - \beta$, between the lines BC and BD. Because the abscissa of point C is $x_A + r\cos\beta - R\cos\alpha = 0$, the value of $R$ is given by

$$R = [x_A + r\cos\beta]/\cos(\theta - \beta). \tag{S1}$$

Define $s$ as the center distance between two adjacent humps. The length of OA is $x_A = (2n-1)s/2$, where $n$ is the contact number between the droplet and cylindrical humps. Then, according to the geometry, the radius of the droplet and the apparent contact angle are given by

$$R = [(2n-1)s/2 + r\cos\beta]/\cos(\theta - \beta), \tag{S2}$$

$$\theta_P = \pi/2 + \theta - \beta. \tag{S3}$$

**The advancing and receding angle asymptotes**

With increasing droplet volume, the common point B slides down along the cylindrical hump. The advancing limit of a droplet is achieved when the liquid-gas interface of the droplet tangentially contacts with the next cylindrical hump, as shown in Fig. S2(b). The radius of the droplet can be obtained in accordance with the geometrical relationships

$$\begin{cases} c^2 = l^2 + [(2n-1)s/2]^2 \\ c^2 = R_a^2 + r^2 - 2R_a r\cos\theta \\ l^2 + [(2n+1)s/2]^2 = (R_a + r)^2 \end{cases}, \tag{S4}$$

where $c$ is the length of line AC, $l$ is the length of line OC and $R_a$ is the radius of the advancing limit of the droplet. The lengths of lines OA and OE are $(2n-1)s/2$ and $(2n+1)s/2$. Then, the droplet radius at advancing limit is calculated as

$$R_a = ns^2/[r(1 + \cos\theta)]. \tag{S5}$$

In real cases, the droplet volume is much larger than the microstructure. Let $n$ be infinite and combine Eqs. (S2) and (S5). The asymptote of advancing contact angle $\theta_a$ can be then obtained as

$$\sin(\pi - \theta_a) = r(1 + \cos\theta)/s. \tag{S6}$$

With decreasing droplet volume, the common point B retreats along the cylindrical hump. The receding limit to the next cylindrical hump involves two conditions. Case (1): The liquid-gas interface tangentially touches the next cylindrical hump, as shown in Fig. S2(c). The geometrical relationship of $R_{r1}$ is then given by

$$\begin{cases} c^2 = l^2 + [(2n-1)s/2]^2 \\ c^2 = R_{r1}^2 + \tau^2 - 2R_{r1}r\cos\theta \\ l^2 + [(2n-3)s/2]^2 = (R_{r1} - r)^2 \end{cases}, \tag{S7}$$

where $R_{r1}$ is the radius of the droplet of the receding limit in case (1), and $(2n-3)s/2$ is the length of line OF. Then, $R_{r1}$ can be obtained as

$$R_{r1} = (n-1)s^2/[r(1-\cos\theta)]. \tag{S8}$$

Similarly, let $n$ approach infinity and combine Eqs. (S2) and (S8). The asymptote of receding contact angle $\theta_{r1}$ in case (1) can be obtained as

$$\sin\theta_{r1} = r(1 - \cos\theta)/s. \tag{S9}$$

Case (2): before the liquid-gas interface contacts the next hump, the common points on the same hump meet, as shown in Fig. S2(d). In this condition, the radius of droplet $R_{r2}$ and the receding contact angle $\theta_{r2}$ are given by

$$R_{r2} = [(2n-1)s/2 + r\cos\beta]/\cos(\theta - \beta), \tag{S10}$$

$$\theta_{r2} = 2\theta - \pi. \tag{S11}$$

Note that the intrinsic contact angle (determined by the chemistry of both surface and liquid and calculated by Young's equation) of the droplet is 60° in Figs. S2(a), (b) and (c), and 120° in Fig. S2(d).

**The droplet volume**

The droplet volume $V_d$ is a function of the intrinsic contact angel $\theta$, apparent contact angle $\theta_p$, contacted humps number $n$, an angle $\beta$ determining the margin common point, hump radius $r$ and center distance between two adjacent humps $s$, as shown in Fig. S3. $V_0$, $V_T$, $V_1$, $V_2$ and $V_3$ are the volumes of the droplet above the BD plane, the trapezoid OABD, the solid line shaded domain, the semicircle of cylindrical hump, and the sector of the outermost cylindrical hump, respectively. Their corresponding expressions are given by

$$V_0 = R^2(\theta_p - \sin\theta_p\cos\theta_p)/2, \tag{S12}$$

$$V_T = [(2n-1)s + r\cos\beta]r\sin\beta/2, \tag{S13}$$

$$V_1 = (2n-1)[r\cos\theta(s - r\sin\theta) + (\theta - \pi/2)r^2]/2, \tag{S14}$$

$$V_2 = (\pi - \beta)r^2/2, \tag{S15}$$

$$V_3 = (n-1)\pi r^2/2. \tag{S16}$$

The total droplet volume $V_d$ is the sum of the above volumes.

$$V_d = \begin{Bmatrix} R^2(\theta_p - \sin\theta_p\cos\theta_p) + [(2n-1)s + r\cos\beta]r\sin\beta \\ +(2n-1)[r\cos\theta(s - r\sin\theta) + (\theta - \pi/2)r^2] \\ -(\pi - \beta)r^2 - (n-1)\pi r^2 \end{Bmatrix} / 2. \qquad (S17)$$

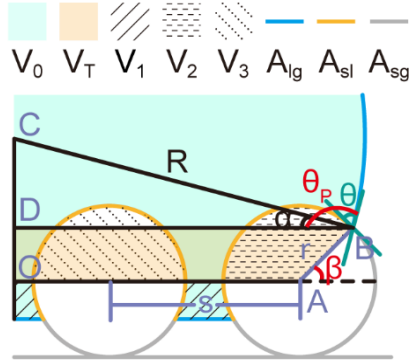

Figure. S3. The droplet volume on a rough surface with cylindrical humps. $V_0$ is the volume of the droplet above the BD plane, and $V_T$, $V_1$, $V_2$ and $V_3$ are the volumes of corresponding domains. $A_{lg}$, $A_{sl}$, and $A_{sg}$ are the interfaces of liquid-gas, solid-liquid and solid-gas, respectively.

**The Gibbs free interfacial energy**

The Gibbs free interfacial energy ($G$) of the whole system can be also calculated as

$$G = A_{sl}(\gamma_{sl} - \gamma_{sg}) + (A_{lg} - A_0)\gamma_{lg}, \qquad (S18)$$

where $\gamma_{sl}$, $\gamma_{sg}$, $\gamma_{lg}$ are the solid-liquid, solid-gas and liquid-gas interfacial tensions. $A_{sl}$ and $A_{lg}$ represent the interfaces of solid-liquid and liquid-gas. $A_{lg}$ consists of two sections, the droplet spherical cap and the plane in microstructures. $A_0$ is the liquid-gas interfacial area as if the droplet is initially cylindrical and not touching the solid surface yet.

$$A_{sl} = 2(n-1)(\pi - \theta)r - (\theta + -3\pi/2)r, \qquad (S19)$$

$$A_{lg} = \theta_P R + (2n-1)(s/2 - r\sin\theta), \qquad (S20)$$

$$A_0 = 2\pi\sqrt{V_d/\pi}. \qquad (S21)$$

Note that the Gibbs free interfacial energy here is the change from the initial state (a separate droplet floating in the air and a dry rough surface) to the present state.

**Numerical simulation**

The theoretical analysis assumes a moving-equilibrium condition, and this section employs a pseudopotential multicomponent lattice Boltzmann method (LBM) to simulate the dynamic jump during advancing and receding [1]. We conducted simulations using Mechsys, and its source code is available at https://mechsys.nongnu.org/. We inject a small amount of water into the droplet step by step and wait until the system reaches equilibrium. The numerical simulation validates our theoretical analysis. Figure S4 shows the comparison between simulated results and analytical analyses (see videos 1 and 2 of advancing and receding processes of simulation). The relationship between contact angle and droplet volume perfectly matches, yet the jump to the next hump is earlier in the simulation (advancing and receding are represented by green and black solid lines in Fig. S4): the jump from the second hump to the third occurs at point B during advancing rather than at the theoretically predicted B′. The actual droplet surface fluctuates during advancing and receding, which could facilitate earlier contact with the next hump. We use numerical tricks such as increasing the droplet viscosity and reducing the density difference between droplet and air to eliminate this effect. Yet this is not the dominant factor. While the fluctuation of the droplet surface could be reduced to about 2 lattices, a numerical experiment pushing a solid hump towards a flat-water surface shows the water is attracted to the hump surface about 7 lattices prior to an actual geometrical contact. This lattice number of 7 is determined by the LBM algorithm. The jump would occur at point B″ if we fully suppress the dynamic droplet fluctuation. A finer mesh pushes the numerical simulation results closer to the analytical solutions. Although this attraction between droplet and hump surface seems to be side-effects of the numerical method, we speculate this could happen in the physical world, for instance when the surfaces are electrically charged.

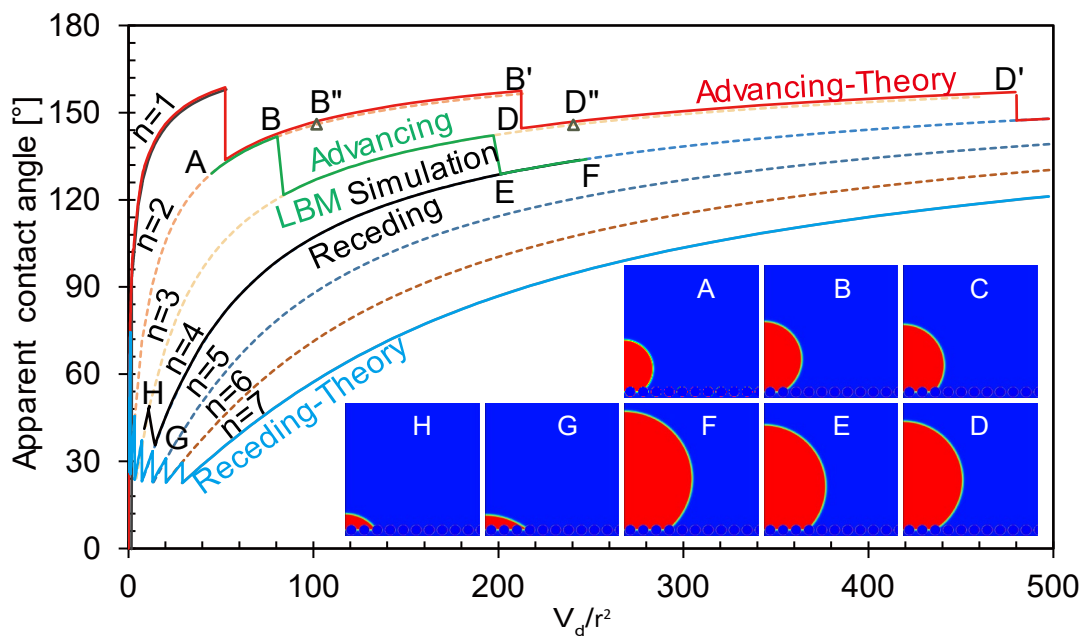

Figure. S4. Apparent contact angles and Gibbs free interfacial energy versus droplet volume with different contacted hump numbers. Note the simulation is conducted with parameters of $\theta$ is 86.5° (this angle corresponds to the case when the inter-molecular interaction strength factor equals 1 in the LBM algorithm [1]) and $s/r = 2.5$.

**Quasi-2D physical experiments**

The intrinsic contact angle of fish wire is measured from 3D CT images (Image_3, pixel size 5.6 μm). A single fish wire was placed approximately vertically in the water. Ten sets of data were measured in five separate slides and averaged to obtain the intrinsic contact angle. Advancing and receding processes on the rough surface (Fig. 4a), made of fishing wires wrapped around designed grooves, were recorded via instrument OCA25. Measured advancing contact angle, 160.4° is averaged by 10 measurements from video_4. The measured receding contact angles range from 26.7° to 65.4°, measured from video_5.

**Video_1:** Simulated advancing process of droplet on rough surface with cylindrical humps.
**Video_2:** Simulated receding process of droplet on rough surface with cylindrical humps.
**Image_3:** CT images of fish wire in contact with water for measuring the intrinsic contact angle of fish wire.
**Video_4:** Droplet advancing on the rough surface made by fish wires wrapped around designed grooves.
**Video_5:** Droplet receding on the rough surface made by fish wires wrapped around designed grooves.
**Video_6:** Top view of droplet receding on the rough surface made by fish wires wrapped around designed grooves.
**Image_7:** CT images of glass beads in contact with water for measuring the intrinsic contact angle of glass beads.
**Video_8:** Droplet advancing on the rough surface made by glass beads.
**Video_9:** Droplet receding on the rough surface made by glass beads.
**Video_10:** Droplet advancing on prewetted rough surface with trapped gas bubbles.
**Video_11:** Shaking of smooth glass slide submerged in water. The water table is 1 mm above the smooth glass slide initially. Shaking the tank induces water receding from the glass surface and exposes the glass surface to air even though the glass surface is lower than the water table. A droplet is left behind when water recedes.
**Video_12:** Shaking of prewetted rough surface submerged in water. The water table is 1 mm above the prewetted rough glass slide initially. The rough surface remains under the water table during tank shaking. This demonstrates that the submerged state, which corresponds to a contact angle of 0°, is at a minimum energy state.
**Video_13:** Shaking of prewetted rough surface with trapped gas bubbles submerged in water. The water table is 1 mm above the prewetted rough glass slide initially. The rough surface remains under the water table during tank shaking.
**Video_14:** Droplet advancing or quick spreading on prewetted diatomite surface.